# Leveraging Technology for Healthcare and Retaining Access to Personal Health Data to Enhance Personal Health and Well-being[1]

Ayan Chatterjee[1*], Ali Shahaab[1], Martin W. Gerdes[2], Santiago Martinez[3], and Pankaj Khatiwada[4]

*[1*,2,4]Department of Information and Communication Technology, Centre for e-Health, University of Agder, 4604 Kristiansand, Norway; {ayan.chatterjee, martin.gerdes, pankaj.khatiwada}@uia.no*

*[1] Cardiff School of Technologies, Cardiff Metropolitan University, CF5 2YB Cardiff, United Kingdom; ashahaab@cardiffmet.ac.uk*

*[3] Department of Health and Nursing Science, Centre for e-Health, University of Agder, 4604 Kristiansand, Norway; santiago.martinez@uia.no*

## Abstract

Health data is a sensitive category of personal data. It might result in a high risk to individual and health information handling rights and opportunities unless there is a palatable defense. Reasonable security standards are needed to protect electronic health records (EHR). All personal data handling needs adequate explanation. Maintaining access to medical data even in the developing world would favor health and well-being across the world. Unfortunately, there are still countries that hinder the portability of medical records. Numerous occurrences have shown that it still takes weeks for the medical data to be ported from one general physician (GP) to another. Cross border portability is nearly impossible due to the lack of technical infrastructure and standardization. We demonstrate the difficulty of the portability of medical records with some example case studies as a collaborative engagement exercise through a data mapping process to describe how different people and datapoints interact and evaluate EHR portability techniques. We then propose a blockchain-based EHR system that allows secure, and cross border sharing of medical data. The ethical, and technical challenges around having such a system have also been discussed in this study.

## 1. Introduction

Securing computerized information and infrastructure has become one of the essential parts of Information Technology in digital assaults. We transfer immense quantities of confidential data electronically and store even more, create real gold mines for hackers who want to steal valuable information following illegal access and denial-of-service attacks. In general, information security assurance is high for – access control techniques, fortification of data over networks, and data security within the enterprise. Secrecy and information assurance are significant global rights.

Effective management of personal and person generated data (PGD) in a highly specialized healthcare system requires digital data sharing to achieve coordinated care. Specific person generated data are very sensitive and are controlled by policies, such as "General Data Protection Regulation (GDPR)" [1], "Normen" [2], "Health Insurance Portability and Accountability Act (HIPAA)" [3]. A signed consent is required before collection of

[1*] Ph.D. Research Scholar.
[1] Ph.D. Research Scholar.

PGD from patients/participants to maintain its security and accessibility. If anyone tries to access it for another purpose, such as curiosity then he/she will be violating general security guidelines [3]. Protected health information (PHI) is a health data linked to information that identifies a "subject"*. "De-identification" [4] techniques are used in eHealth technology to remove the direct or indirect link to re-identify a subject. Health data security policies provide guidelines on how to protect EHR and share it among legitimate users. The sharing must be approved by the legal designates. Patient engagement [5] is an interaction with the health providers and is a key strategy to manage and prevent health risks by accessing their personal and health records (know more about health, feel connected, and act accordingly). However, most of the subjects are concerned about the privacy and misuse of their PHR in digital platforms. Digital health data protection [6] distributes data protection strategies into the following three sub-areas – privacy, security, and trust. Privacy means that people/entities only authorized by the subjects can access electronic health data (EHR). Security protects EHR from external, unauthorized access. Trust ensures authorized sharing of EHR between agreed parties. Security and trust are interrelated. Healthcare specific security standards are depicted in Fig. 1. The following are some well-established methods to ensure health data privacy in digital infrastructure [7][8][9][10][11] - authentication (e.g., TLS, SSL, AuthO), encryption (e.g., RSA, DES, AES), data masking, access control, de-identification, HybrEx, and identity-based anonymization.

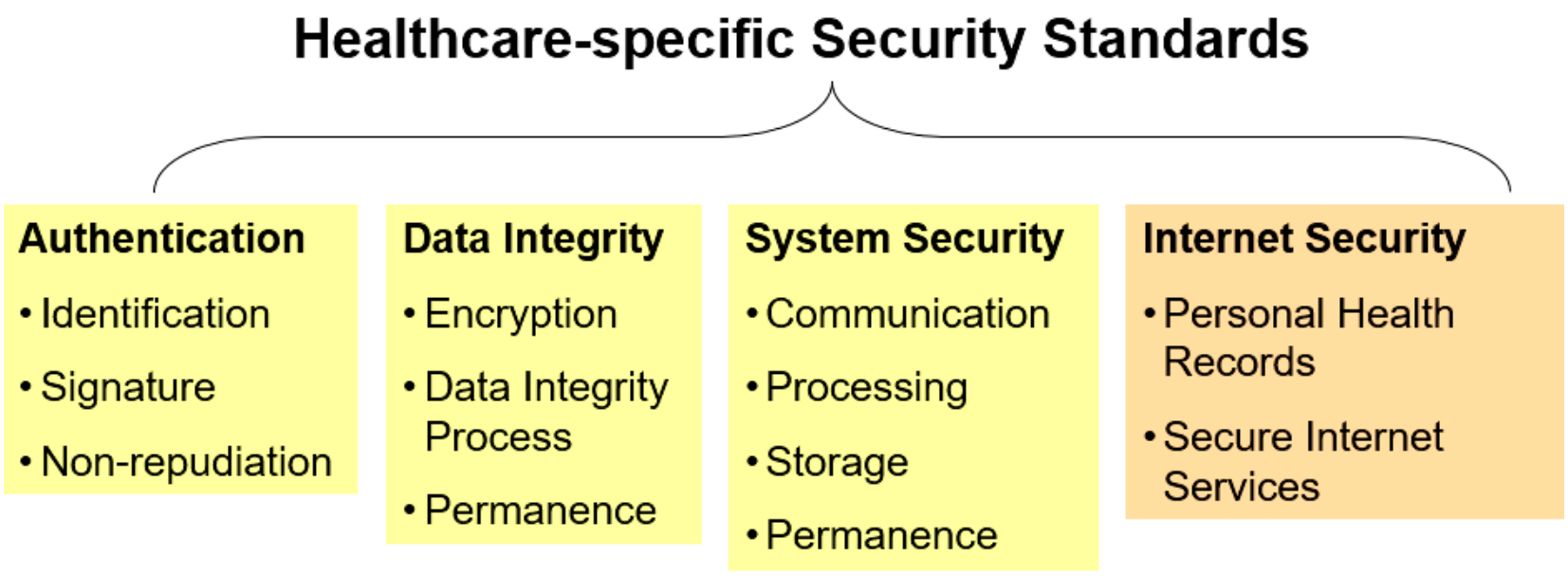

Figure 1: Healthcare specific security standards

Public key infrastructure (PKI) [12] and blockchain [13] are two popular methods for digital data sharing, maintaining the privacy of data. PKI includes the following three components for sharing digital health data among authorized parties – a. public and private keys, b. message/data and digital signature, and c. key organizations (registration authority or RA and certificate authority or CA). The recipient uses the sender's public key to unencrypt the digital signature, verifying that the sender sent it. Once the message is encrypted with the recipient's public key, only they will be able to read it since only they have the matching private key.

## *1.1 Blockchain Technology – a Brief Overview*

Blockchain, the cryptocurrency Bitcoin's underlying technology, is a peer-to-peer, distributed, append-only (P2P) asset transfer network [14]. Blockchain is a type of distributed ledger supported by a group of geographically distributed nodes through different consensus protocols. It democratizes the power through decentralization and prevents information tempering by distributed archiving of information, using multiple encryptions and hashing techniques. Essentially, blockchain is an ordered list of blocks of data where transactions by users are grouped in the form of blocks, and each block has a cryptographic pointer to the

* Subject signifies a patient/participant

** Smart contract is a computer program or a transaction protocol which is aimed to involuntarily execute, control or document legally appropriate events and actions permitting to the terms of a contract or an agreement.

previous block, forming a chain-like structure in which anomalies can be easily detected. Each node supporting the blockchain network maintains a copy of the blockchain and synchronizes it with the rest of the network via different consensus protocols. Unlike distributed databases, the nodes supporting the blockchain network do not inherently trust each other and independently verify every transaction component on the blockchain network, providing a distributed log of events. Since no central authorities manage the blockchain networks, they are highly resistant to censorship and a single point of failure.

Blockchains also allow the users to create smart contracts, which are self-executing contracts containing legal prose of agreements between parties. Since blockchains are distributed, and the data is replicated thousands of times, in most cases, only a commitment or cryptographic proof of the data is added to the blockchain while underlying data is stored in file storage systems such as cloud or interplanetary file system (IPFS) [15]. The data integrity on the blockchain is guaranteed if most nodes in the network are honest. Once a block has been created and appended to the chain, it is computationally costly to change the block's data, and the difficulty increases exponentially as new blocks are appended to the blockchain, making the history immutable. A blockchain system (as depicted in Fig. 2) can be (a) decentralized: a public system where anyone can participate in the network and all records are visible to the public, (b) centralized: a private system where participation is limited and only authorized personnel forms a company can view the records or (c) partially decentralized: consortium among multiple organizations where privileges are managed between multiple organizations. Blockchain technology allows us to record, share, and sync data across geographically distributed parties so that all parties can achieve consensus about the "truth" and the data cannot be altered in the future. This immutability property of the blockchain makes it a suitable candidate for applications requiring accurate history and data sharing among multiple parties, such as medical records and health data [16].

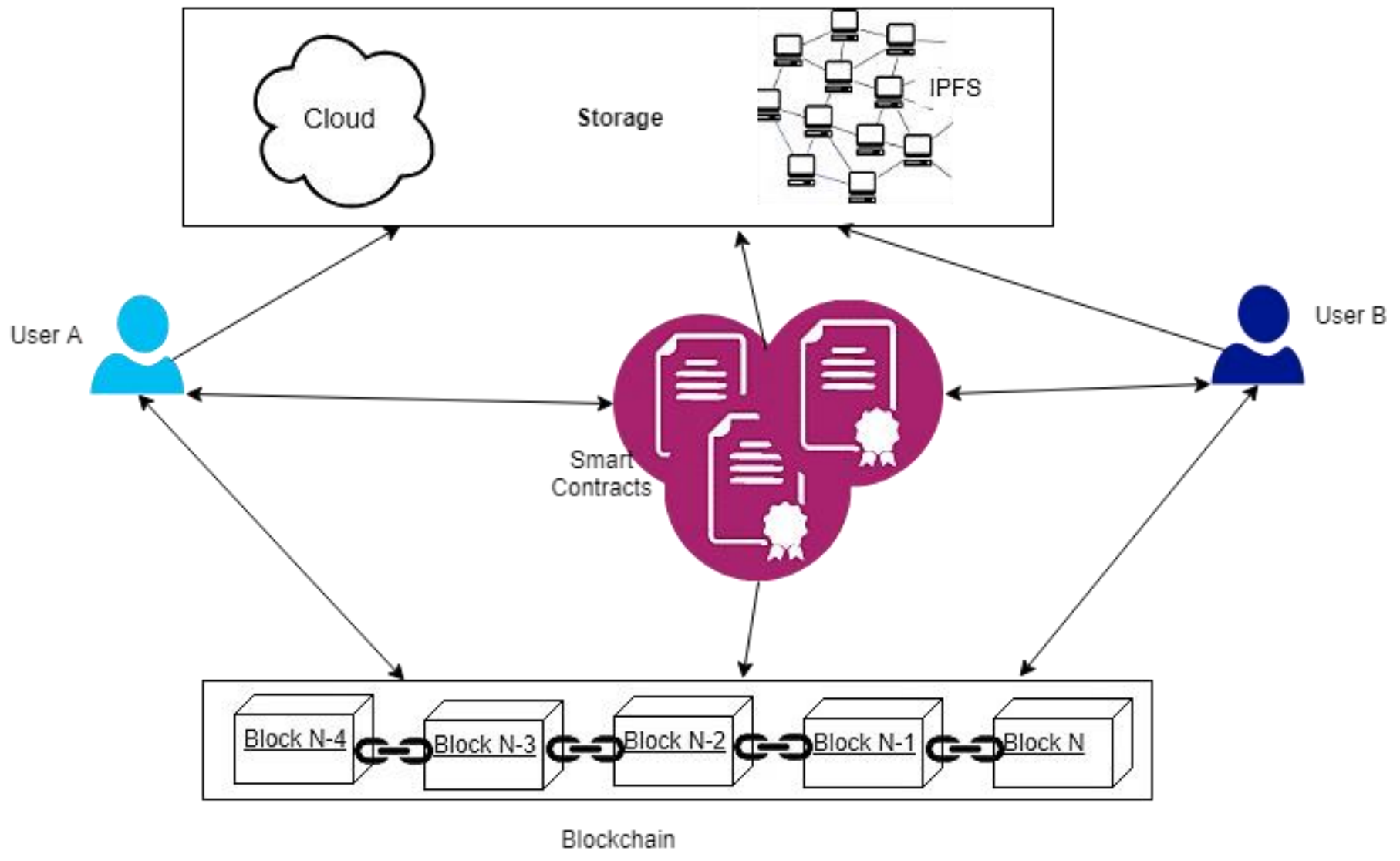

Figure 2: An abstract diagram of a blockchain eco-system. Each block contains cryptographically secured transactions and is appended to the predecessor block. Data is stored in a file storage system and cryptographic proof is posted on the blockchain. Smart contracts can govern the access control and terms and conditions of exchange between user A and user B.

### *1.2 The Work Summary*

The entire study is encouraged from a workshop held at “Not Equal” summer school on digital society for social justice on 27th Aug – 30th Aug 2019, organized by the University of Swansea, UK, in collaboration with other universities and research groups, such as Newcastle University, UK Research and Innovation, University of Sussex, Royal Holloway University, and Engineering and Physical Sciences Research Council (EPSRC), relating to the “algorithmic social justice and digital security” [17][18]. In this four-day workshop, twenty participants were divided into five different groups to explore ways to extend collaborative engagement using creative practices. In the collaborative sessions, participants were asked to write down potential research questions related to automated decision-making, social justice, fairness, and society intersect on day 1. On day 2, groups were formed based on a similar set of research questions. On day-3, LEGO bricks and supporting materials were provided to individual groups to formulate a respective set of research questions and their potential approach for a solution. On day-4, groups were asked to present their views with a group presentation. In the discussion and idea exchange session on day-4, we identified potential challenges related to – a. security algorithm design following ethics, and b. EHR data security, personal access, and portability issues.

In this study, we have focused on the EHR data security. This paper's most important contributions or novelty are as follows – a. *Can we have a global healthcare data portability system with blockchain technology?* and b. *How to ensure the security and privacy of EHR?* The rest of the paper is structured as follows: Section 2 elaborates EHR related challenges in healthcare based on different patient stories. In Section 3, we discuss methods for digital security in healthcare. Section 4 covers a general discussion. The paper is concluded in Section 5. For security and portability challenges of EHR data, we have identified the potential of blockchain technology in Section 3.

## 2. Patient Stories and Identified Challenges

The leading four stories collected from the “Not Equal” summer school workshop [17], are described in the following sub-sections portraying person generated data [18-33] and its security challenges.

### *2.1 Patient Story 1*

Anna Kalb (AK) is a 70-year-old lady who lives in a small Scandinavian village. She widowed two years ago. AK has a daughter who is living approx. 600 kilometers away. She is suffering from a chronic illness. She needs to take care of her health condition. Her general physician (GP) is located far away from her house (approx. 90 mins of drive). Her home is also far away from the nearest hospital. AK has recently been diagnosed with chronic obstructive pulmonary disease (COPD), obesity, and potential diabetes risk. AK also has sporadic episodes of mild depression. Therefore, she needs daily health monitoring and recommendations/suggestion/feedback based on her health conditions. She needs to go through day-to-day analysis of her physiological, and behavioral (activity level, dietary habit, and sleeping pattern) data for personalized recommendation generation, followed by a goal evaluation. AK cannot visit her GP daily. She visits her GP once in a month, and hospital once in a year. AK’s health status will deteriorate steadily unless

* Subject signifies a patient/participant

** Smart contract is a computer program or a transaction protocol which is aimed to involuntarily execute, control or document legally appropriate events and actions permitting to the terms of a contract or an agreement.

monitored well on a timely basis. AK has received a suggestion to assess her health with remote monitoring. She is worried now as she is very new to the digital e-health monitoring system. AK has to learn how to check her health parameters, such as glucose, oxygen saturation level (SPO2), and wear necessary wearable BLE (Bluetooth low energy) devices, and handle them so that data can be appropriately streamed to a centralized decision support system.

Now she is worried about her personal data (EHR) security and its privacy regarding – a. *Who will be the owner of the collected data?* b. *Who will be owning the generated information related to the subjects?* c. *What will be the access control rules to access person generated electronic records?* and d. *Will the research group reuse/ sell the data without any personal consent?*

## *2.2 Patient Story 2*

Ahmed is an 80 years of age male from Pakistan is a case of several billion humans around the world do not have the benefit to have clinical information security and its instant availability. He fled the nation because of strict oppression and guaranteed shelter in the UK. Ahmed's child, who was at that point a UK inhabitant, employed attorneys for his case. Back in Pakistan, Ahmed was determined to have stage 3 cancer, yet the Lawyer in the UK disclosed to Ahmed's child that Ahmed ought not tell about his ailment as it would affect his asylum case. Therefore, Ahmed concealed his complete medical history and treatment plans. Upon reaching the UK, in the asylum, he was diagnosed with health problems and later found that he has cancer. Due to a specific gap in the ongoing treatment, his health condition deteriorated further, and without proper care and medication plan, Ahmed died three weeks later. Ahmed’s life could have been saved if he revealed his medical history after crossing the border with supporting medical documents (EHR).

The story can be narrated with another viewpoint. Ahmed carried all his paper-based medical details with him, but on the way, he lost them. He persuaded the government of his homeland to share his medical history if they could manage a digital copy. However, access to digital health records is blocked outside of the country. Then the question arises – *how could have Ahmed had his medical records protected and ported with him across the border?* He wished his medical records to be secured, portable, and persistent.

## *2.3 Patient Story 3*

Martina is a 40-year-old lady. She has been living in the UK for five years with her little daughter Layla who is six years old. They are registered to the GP at the “XYZ Hospital” where Martina has been working. The GP asked Martina to go for some medical tests as she was suffering from chronic illness and encouraged her about EHR. Besides, she had been asked to wear wearable BLE devices for continuous health monitoring (BP, HR, SPO2, GSR, activities) to sustain a healthy lifestyle. She was worried about the following thoughts – a. *How much has the medical technology developed for personal monitoring health and giving decision support?* and b. *Is her EHR is in the safe hands?*

She consulted again with a GP closest to her home. The GP assured her about the positive aspects of the wearable secured BLE devices for health monitoring (SPO2, heart rate, BP, step count, sedentary bouts, sleeping time, diet planning, GSR response), and explained her about EHR data security. She was delighted. The mom and daughter planned to travel to Hawaii. One beautiful morning, Martina discovered some rashes on Layla's face and hands. The local hospital could not access the EHR of Layla. They tested Layla separately, and she had been found allergic to the pollen. The family is worried now with the portability issues of EHR across the geographical border.

## *2.4 Patient Story 4*

Ruth Jones, who lives in the UK, is 86 years old with chronic health conditions. She used to work as a technologist and a researcher. Ruth needs to consult with his GP on a bi-weekly basis. He has an in-depth interest in technological revolution and ubiquitous health monitoring. He is inclined to digital health monitoring systems to keep himself fit. He uses trusted wearable sensor devices and health applications for daily monitoring of physiological (SPO2, BP, HR, GSR) and behavioral patterns (activity, diet, gait, and sleep) to attain a healthy lifestyle. He is looking for a health management system with personalized recommendation generation capability to manage health goals following health risk prediction. But he is concerned about personal and health data security in a digital system.

He believes that EHR is better than paper-based health record management. He has heard about "Blockchain" technology, a distributed, scalable, ledger-based technology that has a considerable potential to secure personal health data with distributed keys. He is now struggling to know – a. *How does blockchain-based authentication process work?* b. *How to distribute ownership of keys in a blockchain-based authentication process?* c. *Will blockchain help cross-border EHR data portability?* and d. *Will blockchain create any personal burden to manage the assigned key?*

We have identified the following open challenges as defined in Table 1, from all the stories described above.

| Story No. | Challenges |
|---|---|
| 1 | Importance of privacy and protection during handling of personal, and person generated health-related data |
| | Aspects of ethics during electronic monitoring |
| | Potential of a health monitoring system to maintain privacy |
| | Secure portability of EHR |
| 2, 3 | The possibility to make patients as data owner to avoid data exploitation |
| | Leveraging technology to assist the patients in redefining the data ownership |
| | How to make the patients aware of needs and outcomes? |
| 4 | Trust, gender biasness, and ethical aspects of digital healthcare system |
| | Better portability and accessibility of personal health data with blockchain tech. |

Table 1. Identified challenges from the patient stories

# 3. Electronic Health Record (EHR), Its Security and Portability

In this section, we have discussed about the importance of blockchain technology to protect EHR and ensure its secure sharing across international borders.

* Subject signifies a patient/participant

** Smart contract is a computer program or a transaction protocol which is aimed to involuntarily execute, control or document legally appropriate events and actions permitting to the terms of a contract or an agreement.

### 3.1 EHR

It is a systematic collection of patient and population health information in digital format. It helps secure communication of patient's healthcare data between different healthcare professionals, such as GPs, specialists, care teams, and pharmacies. It plays a vital role in Telemedicine where patients and physicians need not be at the same location. Digital health data can be collected from different ways, such as appointments with GPs, hospitals, laboratories, pharmacy, specialists, wearable Bluetooth enabled (BLE) sensors, smartphone applications, self-reported data, digital questionnaires, and feedback forms [34][35]. Data collected from heterogeneous sources are massive, unintuitive, and raw. Therefore, they must be annotated with semantic metadata for more expressive representation, standardization, and creation of rational abstraction. Semantic data [36] helps to create a knowledge base for formal analysis of stored data with reasoning. Reasoning reveals hidden knowledge inside of data either with the rules (semantic web rule language or SWRL [37]) or with machine intelligence (decision support system or DSS [35]). Ontology [38] provides a framework for data interoperability and describes healthcare data collected from heterogeneous sources with proper annotation. DSS analyses healthcare data to generate alerts, reminders, personalized recommendations, and real-time decision aids. HAPI-FHIR [39] is another platform to overcome electronic health data interoperability with JSON [40] annotation. It is a complete implementation of the HL7 FHIR [41] standard for healthcare interoperability.

Healthcare data collected from heterogeneous sources must be stored at a centralized repository in a secure way to protect against illegitimate access, ensuring health data protection, security, data ownership, and privacy. EHR is advantageous than the traditional paper-based medical record-keeping system [35], in terms of time (faster access, ICD ready coding [42], cut down human error and readability issue), cost (reduces labor costs, and operational costs), security (password protection, standardization, encryption, policies, protection against loss and destruction), and eligibility (data ownership, access protection, federal regulations, secure sharing).

### 3.2 EHR Data Sharing Challenges and Opportunities

Jian et al. noticed that no single EHR system provides interoperability since EHR data is generated across different information systems using different schemas [43]. Even though there is a consensus on health data sharing's advantages, addressing challenges around privacy, trust and transparency is vital [43]. An EHR system is only considered "trustworthy" when it can demonstrate the ability to maintain confidentiality, privacy, accuracy, and data security [43]. Different techniques to preserve the traits mentioned above are used in the HER systems. Aleman et al. conducted a systematic literature review on 49 articles and reported encryption techniques (symmetric and asymmetric key schemes) and login/password to be the most common method (13 articles each) of preserving security and privacy of EHR data, followed by PKI based digital signature schemes (11 articles). The role-based access control model was also identified as the preferred choice [44].

Hutchings et al. report a consistent desire to maintain some control over the health data, based on the results from 6859 responses gathered from 35 studies [45]. Some respondents also wanted to know where and when their data was being used, suggesting that access without the patient's consent is a violation of their privacy and concerns about unauthorized access (theft, hacking, or sharing without consent) to the data was also reported [45].

Blockchain technology has been recently recognized as a promising suite of technologies that use some of the best practices in EHR data security and privacy while simultaneously addressing critical data ownership and access concerns.

### *3.3 Blockchain and EHR*

Medrec [46], utilizes a blockchain network to manage and share EMRs and provide an immutable log and access to the medical data owners. No personal data is added to the blockchain. The only hash of the data is stored on the blockchain for data integrity purposes, and "smart contracts" ** are utilized to access the patient's data. Wang and Song proposed a consortium blockchain-based system for the traceability and integrity of cloud-based EHR data, using attribute and identity-based encryption for data security and authorization [47]. Cao et al. identified critical issues with the existing cloud based EHR outsourcing solutions and proposed a public blockchain (Ethereum) integrated EHR management solution for records integrity and suitability [48]. The authors propose recording each operation on the EHR as a transaction on the public blockchain not to be illegally modified in the future, and integrity can be verified. Omar et al. analyze the challenges in centralized EMR solutions and highlighted the potential of blockchain-based solutions [49]. The authors further highlighted implementation challenges in French PHR and provided potential remedies to the highlighted challenges [49]. Zyskind et al. have proposed a general decentralized access control system for personal data. Personal data is kept off-chain with the blockchain's pointers and access control managed by smart contracts [50]. FHIRChain proposed storing metadata of medical data on the blockchain while storing the data off-chain and utilizing smart contracts to exchange data. MedChain proposed a blockchain network, focusing on data sharing between medical stakeholders such as patients, hospitals, and pharmacies [51].

## 4. Discussion

GPs generally have access to the EHR, which is not portable. If a person moves from their own country to another, he/she loses their medical data, and they need to collect lost medical records from the beginning. Here, we have discussed to have a system where people can port their medical data with them when they wish to cross the international border and can interact with the GPs of a new country if the situation demands. The scenario is well depicted in Fig. 3, where the orange barrier prepared with LEGO bricks signifies an international border. Our assessed blockchain-inspired system can securely carry personal medical records or EHR cross-border.

* Subject signifies a patient/participant

** Smart contract is a computer program or a transaction protocol which is aimed to involuntarily execute, control or document legally appropriate events and actions permitting to the terms of a contract or an agreement.

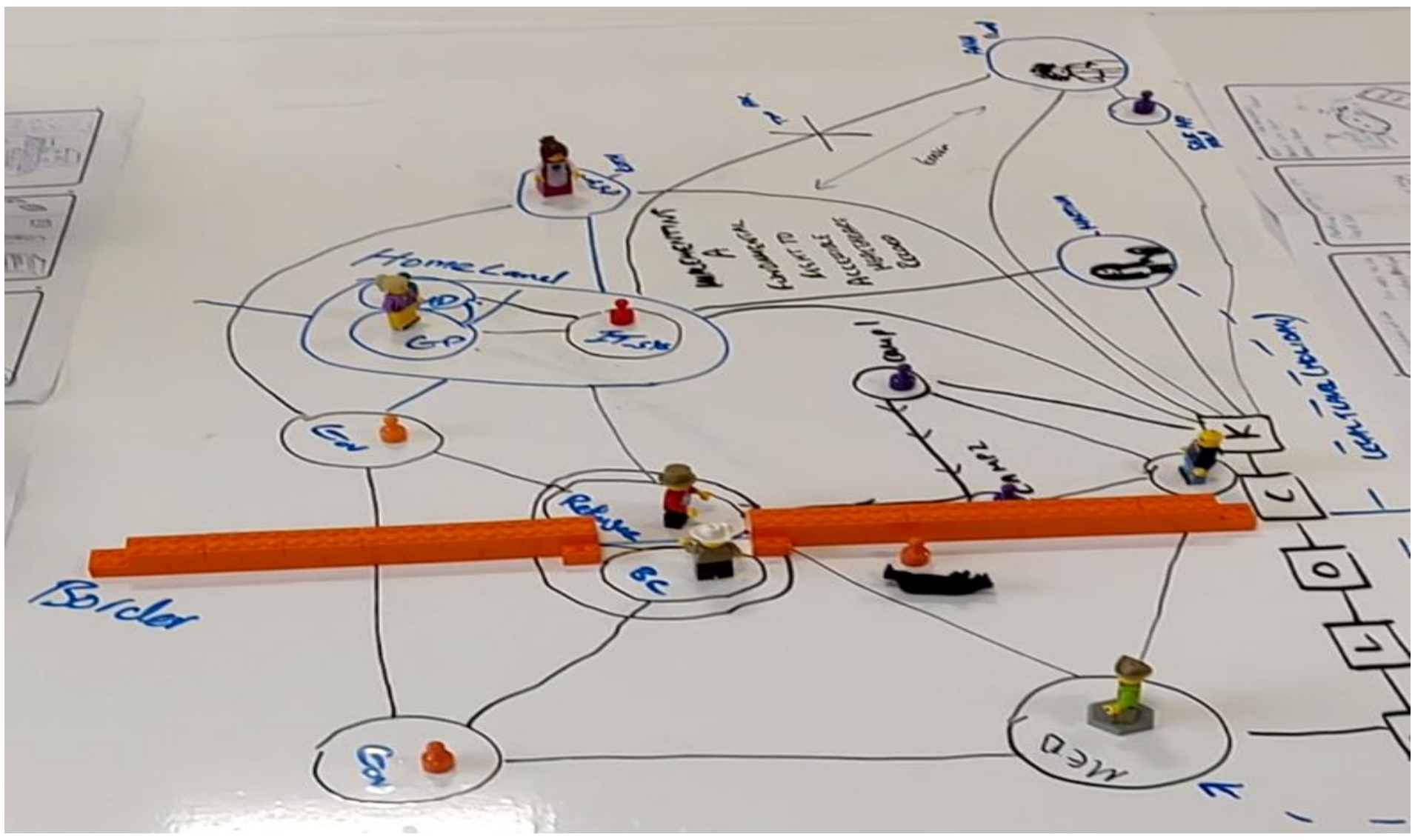

Figure 3. Data portability across the international border

According to a 2018 report of the World Health Organization (WHO), 214 million people have been forcefully displaced and are on the move internationally [52]. When people are forced to leave their homes, they seldomly get a chance to secure their medical records, identities, or other belongings. Norwegian Refugee Council (NRC) found that 70% of Syrian refugees lacked essential identification documents. Having a person's medical history available in a reasonable time can be a matter of life and death, especially for those living in extreme conditions. As a part of the study, health data portability issues for the disaster struck people who are displaced due to socio-political crises, wars, or violence was debated. Specific potential socio-technical solutions as a part of the mind mapping process were considered, and systems' traits were identified with the aim of potentially a global healthcare system, accessible by anyone at any time.

The identified key traits are:

- The solution must be censorship resistant,
- It must hold the integrity of the medical records,
- It must protect the identity of the vulnerable and the patient should have ownership of the data and access control,
- It should allow data aggregation as the displaced person moves from one place to the another,
- Potentially serve as an identity basis for the individual, and the
- Infrastructure should be easy to deploy and maintain.

## *4.1 Censorship Resistance*

A traditional centralized EHR system owned by a state or for-profit organization is inherently prone to censorship. Global politics and macro events can result in the censorship of states or its citizens, giving the monopolies enormous power on the global EHR system. Furthermore, users will not trust a system that is owned by a centralized stakeholder as they will fear the misuse of their data. Blockchain technology's inherent decentralization properties, immutability, auditability, provenance, and availability make it a suitable candidate to act as a backbone for the global health system. Public blockchains, such as Ethereum can be used as a globally available infrastructure to deploy a cryptographically enhanced solution to preserve the EHR data.

### *4.2 Enhanced Integrity and Security*

Medical data of the patient can be encrypted using a Shamir n-1 shared key algorithm [53], where a few stakeholders can participate in the data's decryption. This encrypted data can be hosted on distributed storage such as IPFS or file coin [54]. The hash pointer of the data can be posted on the blockchain to preserve the data's integrity. We recommend using hierarchical keys for transacting on the blockchain so that patient’s complete history can be protected in case of a compromise of a single transaction. Even though the identity on the blockchain is only the public key, users' identities can be further protected by routing the transactions through mixers and using zero-knowledge proofs [55] [56].

### *4.3 Data Aggregation and Identity Basis*

Since the patient owns the hierarchical wallet's root key, the patient will own all the medical history and aggregate medical history. Furthermore, the immutable transactions on the blockchain can prove the refugee’s identity and provide evidence of their journey as they moved from one place to the next.

### *4.4 Ownership and access control*

Blockchain technology ensures data security at storage and helps to a cross-border download of digital medical records safely without disclosing the patient's identity. As part of patient engagement policy, in the blockchain-based EHR security mechanisms, keys are handed over to the trusted parties (such as GPs) and the patient. In the blockchain-based new system, patients must manage their assigned keys to care for their medical data. It might lead to added responsibility or burden to some patients and can make them annoyed.

## 5. Conclusion

Health data is not easily portable. It takes weeks for the data to be ported from one GP to another. Cross border portability is nearly impossible. People from developing nations still rely on carrying their medical records, mostly hard copies. It will be promising if a global healthcare data portability system can be built to port, download, and access a patient's complete medical history legitimately. In our discussed blockchain-based system, patients are encouraged to port their medical records across the international boundaries in a legal, secure, and anonymized way. The study helped us to understand how different data points and agents interact and the importance of EHR cross-border portability.

* Subject signifies a patient/participant

** Smart contract is a computer program or a transaction protocol which is aimed to involuntarily execute, control or document legally appropriate events and actions permitting to the terms of a contract or an agreement.

## Acknowledgements

Thanks to "University of Agder, Department of Information and Communication Technology, Centre for eHealth" for giving me the infrastructure to carry out this study. Additional thanks to Prof. Alan Dix (British author, researcher, and professor at "University of Swansea, UK", specialising in human–computer interaction (HCI)) for organizing the "Not Equal" summer school, 2019.

## References

[1] Voigt P, Von dem Bussche A. The eu general data protection regulation (gdpr). A Practical Guide, 1st Ed., Cham: Springer International Publishing. 2017.

[2] Normen Guidelines. Webpage: https://ehelse.no/normen [accessed on: 9th July, 2020]

[3] Ness RB, Joint Policy Committee. Influence of the HIPAA privacy rule on health research. Jama. 2007 Nov 14;298(18):2164-70.

[4] Neamatullah I, Douglass MM, Li-wei HL, Reisner A, Villarroel M, Long WJ, Szolovits P, Moody GB, Mark RG, Clifford GD. Automated de-identification of free-text medical records. BMC medical informatics and decision making. 2008 Dec 1;8(1):32.

[5] Domecq JP, Prutsky G, Elraiyah T, Wang Z, Nabhan M, Shippee N, Brito JP, Boehmer K, Hasan R, Firwana B, Erwin P. Patient engagement in research: a systematic review. BMC health services research. 2014 Dec;14(1):1-9.

[6] Rumbold JM, Pierscionek B. The effect of the general data protection regulation on medical research. Journal of medical Internet research. 2017;19(2):e47.

[7] Chen CL, Yang TT, Chiang ML, Shih TF. A privacy authentication scheme based on cloud for medical environment. Journal of medical systems. 2014 Nov 1;38(11):143.

[8] Yawn BP, Yawn RA, Geier GR, Xia Z, Jacobsen SJ. The impact of requiring patient authorization for use of data in medical records research. Journal of Family Practice. 1998 Nov 1;47:361-5.

[9] Centers for Disease Control and Prevention. HIPAA privacy rule and public health. Guidance from CDC and the US Department of Health and Human Services. MMWR: Morbidity and mortality weekly report. 2003;52 (Suppl 1) :1-7.

[10] Patil HK, Seshadri R. Big data security and privacy issues in healthcare. In2014 IEEE international congress on big data 2014 Jun 27 (pp. 762-765). IEEE.

[11] Mohammed N, Fung BC, Hung PC, Lee CK. Anonymizing healthcare data: a case study on the blood transfusion service. In Proceedings of the 15th ACM SIGKDD international conference on Knowledge discovery and data mining 2009 Jun 28 (pp. 1285-1294).

[12] Hu J, Chen HH, Hou TW. A hybrid public key infrastructure solution (HPKI) for HIPAA privacy/security regulations. Computer Standards & Interfaces. 2010 Oct 1;32(5-6):274-80.

[13] Ekblaw A, Azaria A, Halamka JD, Lippman A. A Case Study for Blockchain in Healthcare:"MedRec" prototype for electronic health records and medical research data. InProceedings of IEEE open & big data conference 2016 Aug 13 (Vol. 13, p. 13).

[14] Nakamoto S. Bitcoin: A peer-to-peer electronic cash system. Manubot; 2019 Nov 20.

[15] Benet J. Ipfs-content addressed, versioned, p2p file system. arXiv preprint arXiv:1407.3561. 2014 Jul 14.

[16] Shahaab A, Lidgey B, Hewage C, Khan I. Applicability and appropriateness of distributed ledgers consensus protocols in public and private sectors: A systematic review. IEEE Access. 2019 Mar 21;7:43622-36.

[17] Not Equal page. Webpage: https://not-equal.tech/not-equal-summer-school/ (accessed on: 7th July, 2020)

[18] Crivellaro C, Coles-Kemp L, Dix A, Light A. Not-equal: Democratizing research in digital innovation for social justice. interactions. 2019 Feb 22;26(2):70-3

[19] Chatterjee A, Roy UK. PPG Based Heart Rate Algorithm Improvement with Butterworth IIR Filter and Savitzky-Golay FIR Filter. In2018 2nd International Conference on Electronics, Materials Engineering & Nano-Technology (IEMENTech) 2018 May 4 (pp. 1-6). IEEE.

[20] Chatterjee A, Roy UK. Non-Invasive CardioVascular Monitoring. JECET. 2018 Feb;7(1):033-47.

[21] Chatterjee A, Gerdes MW, Martinez SG. Identification of Risk Factors Associated with Obesity and Overweight—A Machine Learning Overview. Sensors. 2020 Jan;20(9):2734.

[22] Chatterjee A, Roy UK. Non-Invasive CardioVascular Monitoring-A Review Article on Latest PPG Signal based on Computer Science Researches. Inter J Res Engine Appl Manag. 2018 Jan;3:1-7.

[23] Chatterjee A, Prinz A. Image Analysis on Fingertip Video To Obtain PPG. Biomedical and Pharmacology Journal. 2018 Dec 25;11(4):1811-27.

[24] Chatterjee A, Roy UK. Non-Invasive Heart State Monitoring an Article on Latest PPG Processing. Biomedical and Pharmacology Journal. 2018 Dec 25;11(4):1885-93.

[25] Chatterjee A, Roy UK. Algorithm To Calculate Heart Rate & Comparison Of Butterworth IIR and Savitzky-Golay FIR Filter. J Comput Sci Syst Biol. 2018 May;11:171-7.

[26] Chatterjee A, Roy UK. Algorithm To Calculate Heart Rate By Removing Touch Errors And Algorithm Analysis. In 2018 International Conference on Circuits and Systems in Digital Enterprise Technology (ICCSDET) 2018 Dec 21 (pp. 1-5). IEEE.

[27] Cretikos MA, Bellomo R, Hillman K, Chen J, Finfer S, Flabouris A. Respiratory rate: the neglected vital sign. Medical Journal of Australia. 2008 Jun;188(11):657-9.

[28] Weiss BD, Mays MZ, Martz W, Castro KM, DeWalt DA, Pignone MP, Mockbee J, Hale FA. Quick assessment of literacy in primary care: the newest vital sign. The Annals of Family Medicine. 2005 Nov 1;3(6):514-22.

* Subject signifies a patient/participant

** Smart contract is a computer program or a transaction protocol which is aimed to involuntarily execute, control or document legally appropriate events and actions permitting to the terms of a contract or an agreement.

[29] Money EW, Caldwell R, Sciarra M, inventors; Life Sensing Instrument Co Inc, assignee. Vital sign remote monitoring device. United States patent US 5,919,141. 1999 Jul 6.

[30] Groff CP, Mulvaney PL, inventors; Groff, Clarence P, Mulvaney, Paul L, assignee. Wearable vital sign monitoring system. United States patent US 6,102,856. 2000 Aug 15.

[31] Dias D, Paulo Silva Cunha J. Wearable health devices—vital sign monitoring, systems and technologies. Sensors. 2018 Aug;18(8):2414.

[32] Menachemi N, Collum TH. Benefits and drawbacks of electronic health record systems. Risk management and healthcare policy. 2011;4:47.

[33] A. Chatterjee, M. W. Gerdes, A. Prinz, S. G. Martinez and A. C. Medin, "Reference Design Model for a Smart e-Coach Recommendation System for Lifestyle Support based on ICT Technologies", in Proceedings of The Twelfth International Conference on eHealth, Telemedicine, and Social Medicine (eTELEMED), pp. 52-58, 2020

[34] Soceanu A, Vasylenko M, Egner A, Muntean T. Managing the privacy and security of ehealth data. In2015 20th International Conference on Control Systems and Computer Science 2015 May 27 (pp. 439-446). IEEE

[35] Chatterjee A, Gerdes MW, Martinez S. eHealth Initiatives for The Promotion of Healthy Lifestyle and Allied Implementation Difficulties. In2019 International Conference on Wireless and Mobile Computing, Networking and Communications (WiMob) 2019 Oct 21 (pp. 1-8). IEEE

[36] Peckham J, Maryanski F. Semantic data models. ACM Computing Surveys (CSUR). 1988 Sep 1;20(3):153-89

[37] Horrocks I, Patel-Schneider PF, Boley H, Tabet S, Grosof B, Dean M. SWRL: A semantic web rule language combining OWL and RuleML. W3C Member submission. 2004 May 21;21(79):1-31

[38] Smith B. Ontology. InThe furniture of the world 2012 Jan 1 (pp. 47-68). Brill Rodopi

[39] Boussadi A, Zapletal E. A fast healthcare interoperability resources (FHIR) layer implemented over i2b2. BMC medical informatics and decision making. 2017 Dec 1;17(1):120

[40] Nurseitov N, Paulson M, Reynolds R, Izurieta C. Comparison of JSON and XML data interchange formats: a case study. Caine. 2009 Nov 4;9:157-62

[41] Bender D, Sartipi K. HL7 FHIR: An Agile and RESTful approach to healthcare information exchange. In Proceedings of the 26th IEEE international symposium on computer-based medical systems 2013 Jun 20 (pp. 326-331). IEEE

[42] Reed GM. Toward ICD-11: Improving the clinical utility of WHO's International Classification of mental disorders. Professional Psychology: Research and Practice. 2010 Dec;41(6):457

[43] Jian WS, Wen HC, Scholl J, Shabbir SA, Lee P, Hsu CY, Li YC. The Taiwanese method for providing patients data from multiple hospital EHR systems. Journal of Biomedical Informatics. 2011 Apr 1;44(2):326-32.

[44] Fernández-Alemán JL, Señor IC, Lozoya PÁ, Toval A. Security and privacy in electronic health records: A systematic literature review. Journal of biomedical informatics. 2013 Jun 1;46(3):541-62.

[45] Hutchings E, Loomes M, Butow P, Boyle FM. A systematic literature review of health consumer attitudes towards secondary use and sharing of health administrative and clinical trial data: a focus on privacy, trust, and transparency. Systematic Reviews. 2020 Dec;9(1):1-41.

[46] Azaria A, Ekblaw A, Vieira T, Lippman A. Medrec: Using blockchain for medical data access and permission management. In2016 2nd International Conference on Open and Big Data (OBD) 2016 Aug 22 (pp. 25-30). IEEE

[47] Wang H, Song Y. Secure cloud-based EHR system using attribute-based cryptosystem and blockchain. Journal of medical systems. 2018 Aug 1;42(8):152

[48] Cao S, Zhang G, Liu P, Zhang X, Neri F. Cloud-assisted secure eHealth systems for tamper-proofing EHR via blockchain. Information Sciences. 2019 Jun 1;485:427-40

[49] El Rifai O, Biotteau M, de Boissezon X, Megdiche I, Ravat F, Teste O. Blockchain-Based Personal Health Records for Patients' Empowerment. InInternational Conference on Research Challenges in Information Science 2020 Sep 23 (pp. 455-471). Springer, Cham

[50] Zyskind G, Nathan O. Decentralizing privacy: Using blockchain to protect personal data. In2015 IEEE Security and Privacy Workshops 2015 May 21 (pp. 180-184). IEEE

[51] Gordon WJ, Landman A. Secure, decentralized, interoperable medication reconciliation using the Blockchain. NIST/ONC. 2016

[52] Kuruvilla S, Bustreo F, Kuo T, Mishra CK, Taylor K, Fogstad H, Gupta GR, Gilmore K, Temmerman M, Thomas J, Rasanathan K. The Global strategy for women's, children's and adolescents' health (2016–2030): a roadmap based on evidence and country experience. Bulletin of the World Health Organization. 2016 May 1;94(5):398

[53] Shamir A. How to share a secret. Communications of the ACM. 1979 Nov 1;22(11):612-3

[54] Benet J, Greco N. Filecoin: A decentralized storage network. Protoc. Labs. 2018:1-36

[55] Hopwood D, Bowe S, Hornby T, Wilcox N. Zcash protocol specification. GitHub: San Francisco, CA, USA. 2016 Oct 4

[56] EY Blockchain · GitHub. Webpage: https://github.com/eyblockchain/. [Accessed: 07-Jul-2020]

* Subject signifies a patient/participant

** Smart contract is a computer program or a transaction protocol which is aimed to involuntarily execute, control or document legally appropriate events and actions permitting to the terms of a contract or an agreement.

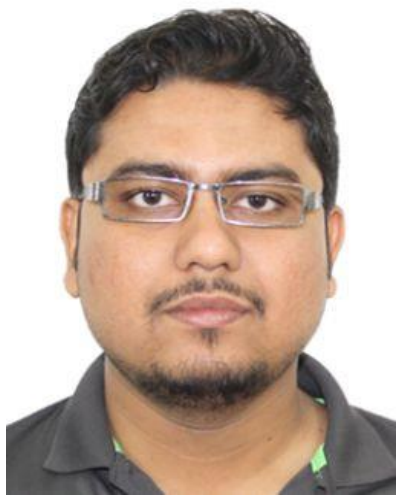

**AYAN CHATTERJEE** received the B.Eng. degree in Computer Science and Engineering from the West Bengal University of Technology, India in 2009, and the master's degree in Information Technology from Jadavpur University, India in 2016. He worked as an Associate Consultant in Tata Consultancy Services ltd., India from 2009 to 2019 and was deputed to Denmark and the Netherlands for 3.4 years as a Java Technical Lead, Application Designer, and Data Analyst. He is now pursuing his Ph.D. in the department of Information and Communication Technologies, University of Agder, Norway since 2019 with a specialization in eHealth. His research interests are machine learning, deep learning, statistical analysis, persuasive computing, domain ontology, human centered design, and data mining.

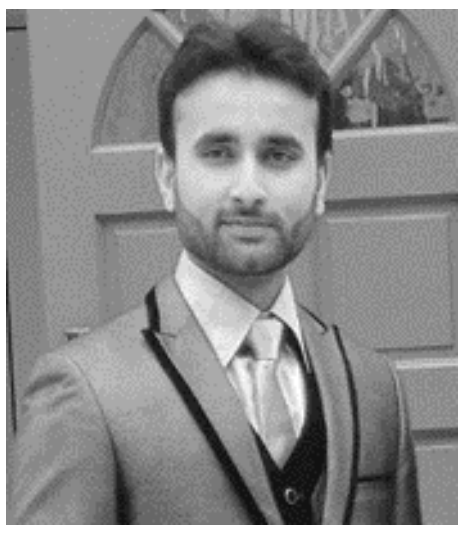

**A. SHAHAAB** is a PhD candidate at Cardiff Metropolitan University. His research focuses on the feasibility of Distributed Ledgers (DLTs) and Blockchain Technology to guarantee the integrity of Companies House UK data. Key areas of his research are DLTs frameworks, security, privacy, and immutability aspects of DLTs and Blockchain as well as legislative and deployment challenges. He also has a passionate interest in utilising blockchain technology to solve real world issues around trust, transparency, and identity. Having MSc in advanced computer science from University of St. Andrews, UK and BS in Electrical and Computer Engineering from Comsats Islamabad, Pakistan, Ali has a strong grip on software systems, distributed computing, cryptography, and information systems.

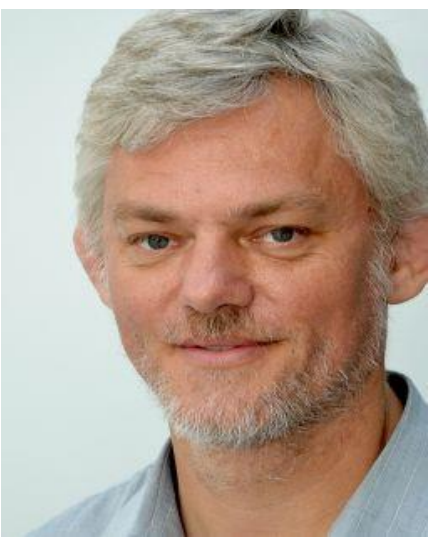

**Martin W. Gerdes** received Diploma in Engineering (equivalent to M.Sc.) in General Electrical Engineering, RWTH (Technical University) Aachen, in 1998, emphasis on Information and Communication Technologies. He completed Ph.D. in Information and Communication Technologies, with specialization in eHealth; University of Agder (UiA), Grimstad, Title: "Holistic System Design for Distributed National eHealth Services". He is employed at the same university as an Associate Professor in the department of information and communication technology with teaching and supervision experience since 2014. His main research interest lies within smart, safe, reliable, and user-friendly eHealth solutions for remote monitoring and patient support, and AI technologies for decision support systems, utilizing Information and Communication Technologies (ICT) and Internet of Things (IoT). Dr. Gerdes is supervising Ayan Chatterjee in his doctoral thesis as a principal supervisor. Before coming to academics, he served Ericsson Eurolab for more than a decade.

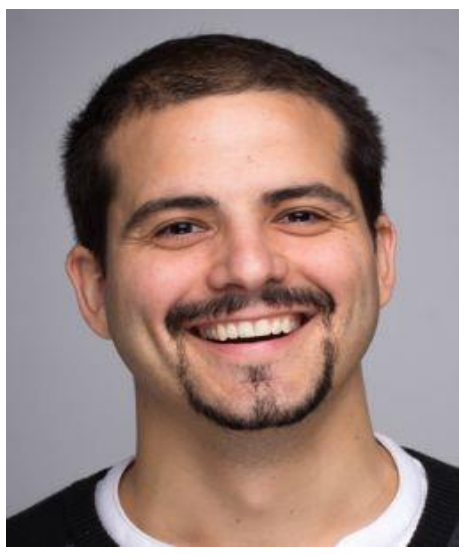 **Santiago Gil Martinez** was awarded the Alison Armstrong Research Study Place to do his PhD in Human-Computer-Interaction at the University of Abertay, UK, in 2009 where he worked in a multidisciplinary environment with psychologists, sociologists and health professionals. His research focuses on user-centered design and usability. Dr. Martinez's background is Computing Engineering, and he worked in Human-Computer Interaction (HCI) for the last nine years. He is interested in special users, such as the elderly, infants, those with disabilities and those without the Information and Telecommunications Technology (ICT) experience. He has developed specific methods for his research, working with a wide range of those established as Participatory Design, user-centered design and ethnographic. He is now employed in University of Agder, at the department of Health and Nursing Science as an Associate Professor. He has a good research collaboration with Scotland, UK. Dr. Martinez is supervising Ayan Chatterjee in his doctoral thesis as a co-supervisor.

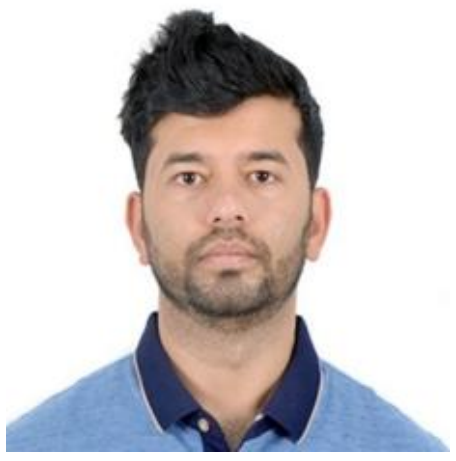 Pankaj Khatiwada received his B.Eng. degree in Electronics and Communication Engineering from Kathmandu Engineering College, Nepal in 2013, and M.Eng. in Information and Communication Technology from the University of Agder, Norway, in 2019. He worked as Project Engineer in Arya Nirman Sewa, Nepal, from 2013 to 2015 and responsible for handling different electrical and communications technologies projects. He is now working as the Researcher in the eHealth department at the University of Agder, Norway, since Jan 2020 with a specialization in eHealth. His research interests are health infrastructure, health data security, AI in healthcare, and blockchain in health data.

* Subject signifies a patient/participant

** Smart contract is a computer program or a transaction protocol which is aimed to involuntarily execute, control or document legally appropriate events and actions permitting to the terms of a contract or an agreement.